# Dominating Attributes Of Professed Firm Culture Of Holding Companies – Members Of The Bulgarian Industrial Capital Association

**Kiril Dimitrov\*, Marin Geshkov\*\***

## Summary

This article aims to outline the diversity of cultural phenomena that occur at organizational level, emphasizing the place and role of the key attributes of professed firm culture for the survival and successful development of big business organizations. The holding companies, members of the Bulgarian Industrial Capital Association, are chosen as a survey object as the mightiest driving engines of the local economy. That is why their emergence and development in the transition period is monitored and analyzed. Based on an empirical study of relevant website content, important implications about dominating attributes of professed firm culture on them are found and several useful recommendations to their senior management are made.

**Keywords:** organizational culture, firm culture, corporate culture, professed culture
**JEL:** L20, M14.

## Introduction

Globalization and modern ways of collaboration among target business entities are the main preconditions for the increasing fluidity in the nature of culture, driven by crossing continental, national and regional borders, co-mingling, hybridizing, morphing and clashing among the numerous manifestations of its attributes. On the one hand, media, migration, telecommunications, international trade, information technology, supranational organizations, small or medium-sized companies, trying to internationalize their business activities, the realizations of any types of public-private partnerships, and unfortunately terrorism may be defined as some of the most important playgrounds for such encounters or even clashes, realized through diverse nuances in the meaning, embedded in the respective cultural attributes along their comparative positioning on the continuum "integration – collision". On the other hand, dissolution of organizational borders, deliberately activated by contemporary managers through numerous forms of employee employment, interaction and engagement, applied to different personnel categories, even sometimes oriented to groups, officially registered outside the company, widens the scope and range of performed HRM activities, designed not only for permanently full-time employees, but also for permanent part-time staff, fixed-term, on-call labour, leased, distant or potential employees, and in some industries and cases even to employees of company constituencies, irrespective of

\* University of National and World Economy, Department of Industrial business

\*\* University of National and World Economy, Department of Industrial business

their physical work locations (see Nakata, 2009; Ulrich, Younger, Brockbank, Ulrich, 2012; Armstrong, 2012). In such realm the repeated processes of effective and efficient formulation and implementation of leadership intentions for organizational development and predictability of behavior for partner organizations come of great importance. For this reason the attributes of professed organizational culture attract the attention of different social actors (business owners, managers, employees, opinion leaders, politicians, etc.), demonstrating different interests in the survival and adaptability of a target entity and the integration among the members of its personnel. That is why the current article is aimed at outlining and analyzing the place, role and dominating content of professed firm culture in the group of big holding companies from the Bulgarian Industrial Capital Association within the virtual realm (***, 2017g). This aim may be decomposed to three main tasks, as follows:

● Disclosing the importance of the cultural phenomena in the business world and their impact within the organizational settings, paying exclusive attention to the results from leadership efforts, oriented at company survival and prosperity, and incarnated in the formulation of key attributes of the professed firm culture, publicly and bravely announced on the internet.

● Describing important methodological steps for the respective empirical survey of a selected group of local holding companies.

● Analyzing and summarizing the results from the performed content survey of target company websites.

That is why appropriate research methods are applied – observation of key attributes of company internet sites, analysis, synthesis, induction and deduction, critical analysis, expert appraisal and statistical methods (one-dimensional distributions of target variables and important cross tabulations among some of them).

## 1. Diversity of cultural phenomena at organizational level

Different approaches in defining the essence of organizational culture may be applied as useful means for identifying the place and significance of professed corporate culture among all cultural manifestations within organizational settings, i.e. content approach, problematic approach, functional approach, etymological approach and important research streams, influencing how the concept of organizational culture is perceived. Viewing the cultural phenomena through different perspectives allows social actors (researchers, managers, opinion leaders, etc.) to use a number of constructs instead of "forms" as follows: elements, components, categories of the organizational culture. A detailed list of organizational culture forms, arranged by degree of visibility and levels of consciousness in the perceptions of their meanings for the unbiased observer, is shown in figure 1.

**Fig. 1.** *Forms of organizational culture*

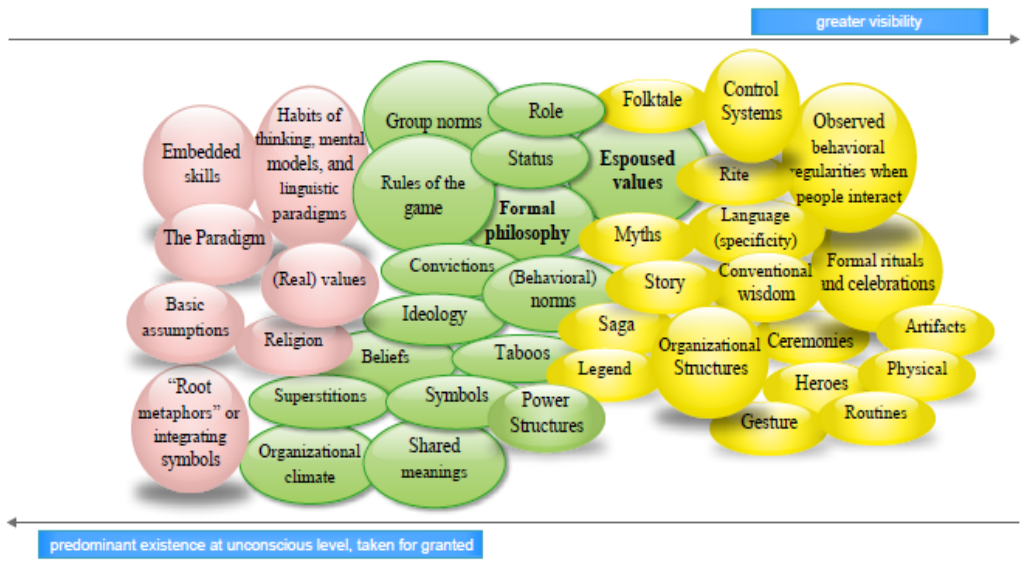

Sources: Johnson, Scholes, Whittington (2008); Schein (2004); Schein (2010); Paunov (2005); Trice, Beyer (1993); Paunov (2015).

In this arrangement the forms are inherently classified in different colors, as if filling with content and belonging to Edgar Schein's model of organizational culture levels (Schein, 2010, p24) – with yellow - representing the most cursory cultural level; velvet color- representing the deepest cultural realizations; and green - representing the cultural attributes in-between, i.e. espoused beliefs and values. The last mentioned layer is the place where the elements of professed organizational culture may be observed. For the purpose of this analysis researchers' interest in this layer is strictly restricted to norms that acquired civilization statute through the creation and official adoption of company documents in the sphere of espoused values, formal philosophy, ideology, behavioral norms, constituting the first step in the strategic management process usually undertaken by senior leaders in business organizations. Furthermore, some nuances in the meanings of the presented forms may overlap, because of different interests demonstrated by the respective constituencies (for example anthropologists, psychologists, entrepreneurs, etc.) which may contribute to some instability and ambiguousness in generated survey results.

The concentrated perspective on the problematic approach to defining the meaning of organizational culture may also serve as a means to disclose important nuances in its officially proclaimed characteristics while presenting the solution stages for the two main issues, i.e. (1) survival and adaptation of the group to the external environment and (2) internal integration among the members in the group[1] (see table 1). Through the lens of the first organizational culture issue some characteristics of the professed culture may be detected as the company mission, the organizational purpose, the manifest functions and some important succinctly described characteristics of the general overall company development strategy. The second organizational culture issue outlined gives a hint to the existence of the official organizational ethics, revealed by the enacted ethical code, dress code and ideology. But as a rule a great deal of steps and related tasks that group members have to perform in order to solve pending and emerging cultural issues, representing inseparable part of confronted business related problems, remain unofficial and even exist at the human unconscious level, summarized as "the way we do things around here" (Bower, 1966; Schein, 1968, 1978; Van Maanen, 1976, 1979b; Ritti and Funkhouser, 1987, Deal, Kennedy, 1982).

---

[1] In the cultural studies literature the term "group" is frequently used, substituting the widely applied term in managerial literature "team". Here the authors use these two constructs as synonyms, but from the perspective of business management it should be pointed out that a team means a structured group or community of persons who make efforts, cooperate systematically, share common resources, organized for a particular purpose, for example to conduct business. While the construct 'group' may be defined as all the people, waiting at terminal 2 of Sofia airport.

**Table 1.** *Comparing the steps for solving the two main cultural problems for the group members*

| PROBLEMS | | | | | |
|---|---|---|---|---|---|
| (1) SURVIVAL AND EXTERNAL ADAPTATION | | | (2) INTERNAL INTEGRATION AMONG THE GROUP MEMBERS | | |
| Steps | | Description | Steps | | Description |
| 1.1. | Mission and Strategy | Obtaining a shared understanding of core mission, primary task, and manifest and latent functions. | 2.1. | Creating a common language and conceptual categories. | If members cannot communicate with and understand each other, a group is impossible by definition. |
| 1.2. | Goals | Developing consensus on goals, as derived from the core mission. | 2.2. | Defining group boundaries and criteria for inclusion and exclusion. | The group must be able to define itself. Who is in and who is out, and by what criteria does one determine membership? |
| 1.3. | Means | Developing consensus on the means to be used to attain the goals, such as the organization structure, division of labor, reward system, and authority system. | 2.3. | Distributing power and status. | Every group must work out its pecking order, its criteria and rules for how members get, maintain, and lose power. Consensus in this area is crucial to helping members manage feelings of anxiety and aggression. |
| 1.4. | Measurement | Developing consensus on the criteria to be used in measuring how well the group is doing in fulfilling its goals, such as the information and control system. This step also involves the cycle of obtaining information, getting that information to the right place within the organization, and digesting it so that appropriate corrective action can be taken. | 2.4. | Developing norms of intimacy, friendship, and love. | Every group must work out its rules of the game for peer relationships, for relationships between the sexes, and for the manner in which openness and intimacy are to be handled in the context of managing the organization's tasks. Consensus in this area is crucial to help members manage feelings of affection and love. |
| 1.5. | Correction | Developing consensus on the appropriate remedial or repair strategies to be used if goals are not being met. | 2.5. | Defining and allocating rewards and punishments. | Every group must know what its heroic and sinful behaviors are and must achieve consensus on what is a reward and what is a punishment. |
| | | | 2.6. | Explaining the unexplainable -ideology and religion. | Every group, like every society, faces unexplainable events that must be given meaning so that members can respond to them and avoid the anxiety of dealing with the unexplainable and uncontrollable. |

*Source: Schein (2010, 2004, 1990)*

The functional approach to defining organizational culture may also be related to professed organizational culture through the content of some of the functions, especially when a detailed list of functions (Paunov, 2015, 2005) is chosen for the purpose of this analysis, because most of the reviewed functions are oriented to penetrating into deeper, hidden and unwritten cultural characteristics (see table 2). Thus, it becomes evident through the provided descriptions that at least two organizational culture functions, i.e. socialization function and identification function, are directly related to the attributes of professed company culture - motto, mission, vision and basic characteristics of corporate strategy, business-unit strategy or functional strategies. The potential existence of some indirect relationships between professed organizational culture elements and the other functions from the list may also be implied, based on the elaboration of managerial technology in the company, concerning different spheres of performed organizational activities.

**Table 2.** *The functional approach in defining organizational culture*

| PROPOSED LIST OF FUNCTIONS | DESCRIPTIONS OF THE FUNCTIONS |
|---|---|
| 1. It assists in the formation of a system (formative function) | It ensures unanimity in attitude and action of a group of people. |
| 2. Socialization function | Human sense of belonging, need for predictability, security, comfort and realization (self-actualization) cannot be satisfied without internalizing or taking organizational values as one's own or supplying individual's values to the organization. |
| 3. Communicative function | It provides a shared meaning of everything that happens in an organization for its members, facilitates the coordinated collaborative actions among its members. |
| 4. Identification function | Culture, expressed by the motto, mission, and vision or incarnated in corporate strategy, business-unit strategy or functional strategies, may be viewed as a differentiator of the definite organization in comparison to the other companies in a definite sector or the entire economy, providing special character and uniqueness of the entity. If strategy and culture reinforce each other, employees find it natural to be committed to the strategy. |
| 5. Integrative function | Gives its members common organizational identity, i.e. generates the sense of "… us" for the people in the organization or closing organization's ranks, thus sustaining and developing the team spirit and commitment. |
| 6. Instructive (educational) function | The process of generating common organizational identity for the entity's members is always achieved by overcoming (yielding) a part of the individual's ego. |
| 7. Adaptation function | Organizational culture supports one-way, synchronized, logically successive actions of the organization's members who in this way percept and react to all kinds of outside events. |
| 8. "Power/roles" function | The organizational culture creates and sustains the necessary distribution and balance of power and influence among positions, informal roles and definite incumbents. It grades by relative importance in a hierarchy all the positions, roles and individuals in the organization, thus facilitating the communications. |
| 9. Genetic function | The organizational culture preserves in time the unique parameters of the entity, ensuring continuity and heredity in the organization from generation to generation. |
| 10. Regulating function | It mitigates strong emotions and dangerous behaviors (for example: fear, frustration, aggressiveness, fury or excessive optimism). |

*Source: (Paunov, 2015, 2005).*

The etymological approach in defining the exact term of "organizational culture" may also reveal important nuances in the relationship between the elements of professed culture and the holistic cultural context for the company, bearing in mind the availability of a number of diverse definitions for the aforementioned term. For the purpose of the current analysis the authors' attention is restricted to just three very popular definitions (see table 3). As a rule, the majority of the researchers in the business-related cultural field are interested predominantly in tacit, hidden and taken-for-granted cultural attributes, thus inherently neglecting the study of officially proclaimed cultural attributes and leaving them as a priority of other inter-disciplinary sciences, for example strategic management where not all of them receive balanced attention by senior managers and researchers.

**Table 3.** *Selected definitions of organizational culture*

| AUTHOR | DEFINITION |
|---|---|
| *Edgar Schein* (2010, p.18)<br>• In this definition the author applies the term group culture, but sometimes the group may represent the whole personnel of a target company. In other cases several groups may constitute the entire personnel of a business entity. The concept of professed culture is just implied by introducing the teaching process for the new employees. | Organizational culture is "a pattern of shared basic assumptions that was learned by a group as it solved its problems of external adaptation and internal integration that has worked well enough to be considered valid and, therefore, to be taught to new members as the correct way to perceive, think, and feel in relation to those problems." |
| *Andrew Pettigrew* (1979)<br>• Ideology as an element of professed organizational culture is mentioed here in a semantic part of the definition. | "Organization culture is the system of common and jointly perceived meanings, valid for a certain group towards a given moment and it is the source of symbols, language, ideology, beliefs, rituals and myths in the organization". |
| *Marin Paunov* (2015, 2005)<br>• The socio-anthropological perspective is accepted here with a great business humor emphasis. The concept of professed culture is just implied by defining the scope of organizational culture. | "Organizational culture is everything in an organization (even if the boss is not cultural), because the entity comes into existence by people, exists for and through its people." |

The main research streams within organizational culture studies may also be applied as indirect means of identifying key elements of the professed company culture (see table 4), because the three main questions that may find appropriate answers here are:

• What are the elements of professed organizational culture for a target entity that may be detected publicly on paper or on the internet?

• To what extent do the structure and content of the identified proclaimed organizational culture elements correspond to the best practices in making the "first step" in strategic management by senior executives, described in leading scientific literature?

• Are there any or what are the differences between the officially declared cultural characteristics for a target business organization by senior management and the everyday in-company life, directed by a unique set of basic assumptions?

**Table 4.** *Basic research streams, influencing how the concept of organizational culture is perceived*

| RESEARCH STREAM | DESCRIPTION | DISADVANTAGES |
|---|---|---|
| Survey Research method | From this perspective, culture has been viewed as a property of groups that can be measured by questionnaires leading to Likert-type profiles. | 1. This approach assumes possessing knowledge of the relevant dimensions to be studied. 2. Nevertheless the use of large samples, strong doubts arise in relation with broadness and relevancy of the aforementioned dimensions (initial item set) in order to capture critical cultural themes for a target business organization. 3. It remains unclear whether something as abstract as culture can be measured with survey instruments at all. |
| Analytical Descriptive method | In this type of research, culture is viewed as a concept for which empirical measures must be developed, even if that means breaking down the concept into smaller units so that it can be analyzed and measured. Thus organizational stories, rituals and rites, symbolic manifestations, and other cultural elements come to be taken as valid surrogates for the cultural whole. | This approach fractionates a concept whose primary theoretical utility is in drawing attention to the holistic aspect of group and organizational phenomena. |
| Ethnographic method | In this approach, concepts and methods developed in sociology and anthropology are applied to the study of organizations in order to illuminate descriptively, and thus provide a richer understanding of certain organizational phenomena that had previously not been documented fully enough. Better theory is built through this approach. | 1. This approach is time consuming and expensive. 2. Many more cases are needed before generalizations can be made across various types of organizations. |
| Historical method | Though historians have rarely applied the concept of culture in their work, it is clearly viewed as a legitimate aspect of an organization to be analyzed along with other factors. | 1. It possesses drawbacks similar to those pointed out for the ethnographic approach. 2. The insights that historical and longitudinal analyses can provide often offset the aforementioned drawbacks. |
| Clinical Descriptive method | Organizational phenomena are observed by consultants as a byproduct of their client services, although preliminary defined by the client domain of respective observation. In this way penetration into the higher levels of management is ensured where organizational policies, reward and control systems are designed. The empirical knowledge gained from such observations provides a much needed balance to the data obtained by other methods because cultural origins and dynamics can sometimes be observed only in the power centers where elements of the culture are created and changed by founders, leaders, and powerful managers. | This method provides neither the descriptive breadth of an ethnography nor the methodological rigor of quantitative hypothesis testing. |

*Source: Schein (1990)*

The search for veracious answers to the first two questions does not require a compulsory penetration through organizational boundaries and dominating in-company climate. But finding a true answer to the third question may not proceed without considering the opinions of insiders. That is why it may be concluded that at the current evolutionary stage of the field, a combination of ethnographic and clinical research seems to be the most appropriate basis for not only trying to understand the concept of culture, but also revealing its officially declared part to the constituencies.

No matter what approach to defining organizational culture construct is used, the importance of professed company culture is unquestionable, because it represents the main tool, guiding the desired future development of the entity, set by senior managers and communicating its public image to important constituencies, even in the internet realm.

## 2. Outlining basic attributes of professed culture in business organizations

Relying on Keenoy's (1999) hologram perspective makes it possible for scientists to assume the official firm documents used to communicate the professed culture of a target entity as a coherent, comparatively stable, but amorphous whole of multifaceted and interrelated norms, accepting civilization statute in numerous forms (e.g. official company documents) – mission, vision, motto, credo, corporate/ firm/ official philosophy/ policy, firm/ our values, company history, information "about us", code of conduct/ ethical code, corporate/ firm principles, purpose, firm/ corporate/ organization culture, corporate social responsibility/ sustainability, slogan and manifesto (see Dimitrov, Ivanov, Geshkov, 2016, 2015). In this way the specific necessities of separate entities are efficiently considered from at least three perspectives – cultural, strategic and communication. From that array the company mission seems to be the most popular attribute of professed organizational culture among the communities of practicing managers and researchers that may be easily checked by undertaking deliberate searches through key words, representing the aforementioned official company documents in selected scientific databases, providing scientific articles, conference proceedings and book chapters (2017a, 2017b, 2017c, 2017d, 2017e, 2017f). A succinct definition for this construct (Zlatev, 1999) presents it as a specific purpose that possesses basic characteristics; is applied as a device to define dominating attitudes to key constituencies; serves as a means of proclaiming considerations (underlying reasons) for survival and successful development of the business organization, formulated by senior management; requires performing a thorough analysis of certain factors; may be specified through key adjectives. This semantic structure is presented by figure 2. Furthermore, the relationship between close constructs, describing important attributes of professed culture, is indirectly outlined here, i.e. through the comparison between mission statement and vision. The last one is characterized by a narrower span for shades of meaning, included in it, i.e. it constitutes just one factor for analysis in a series of five ones.

**Fig. 2.** *Basic structural view of company mission*

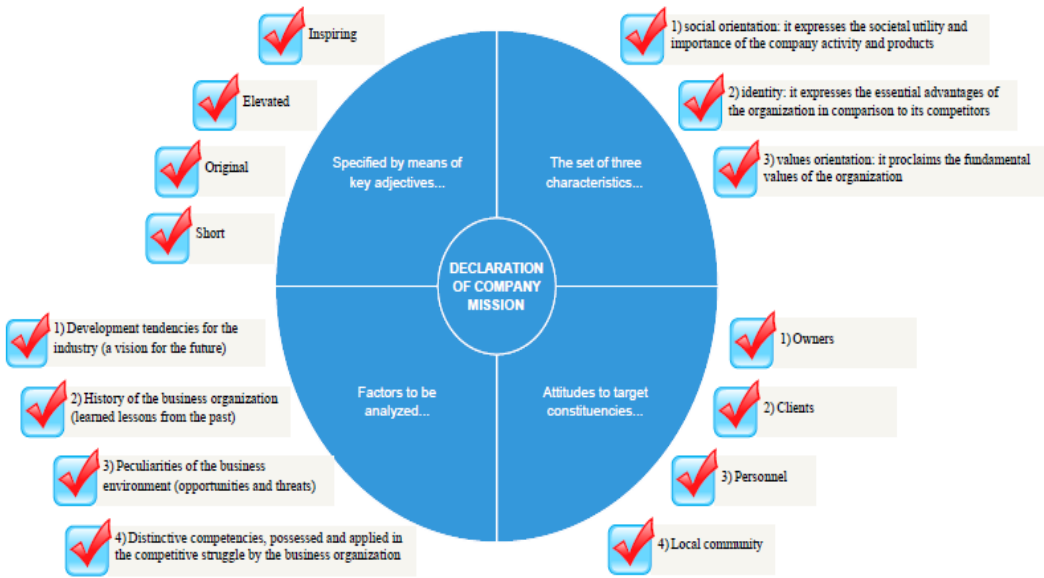

The great importance and richness in the outlined nuances of the meaning of the term mission statement for the survival and sustainable competitive success of business entities may be detected by differentiating approaches, applied to its surveying and expressed uncertainty by researchers and managers, as follows:

● Observed variances in its insertion at the initial stage in the strategic management process, realized in business organizations – management by objectives (Drucker, 1971), strategic management theory (Pearce, 1982), strategic planning (David, 1989).

● Its potential application as an efficient managerial instrument to outline the specific scope of the firm's activities, i.e. manufactured products, delivered services, occupied markets, and deliberate differentiation of a target company without necessarily pointing out its advantages in comparison to the competitors (David, 1993; Sufi, Lyons, 2003; Strong, 1997).

● Observed differences in the scope of the set of embedded specific aims the business organization pursues through the mission (see table 5) (Bartkus, Glassman, McAfee, 2000; Bart, Bontis, Taggar, 2001; Bart, 1997; Sanchez, Heene, 2004; Desmidt, Princie, Decramer, 2011).

**Table 5.** *The scope of embedded specific aims, pursued by the company through formulation of a mission statement*

| Oriented inwards… | Oriented outwards… |
|---|---|
| To increase the level of congruence (connectivity) in the business organization through provision of direction and purpose. | |
| To serve as a control mechanism upon the employees. | |
| To serve as an organizational decision-making guide, based on shared values and established behavioral standards. | |
| To lend significance to respective job activities through inspiration and motivation of personnel members. | |
| To serve as a means for obligatory taking into consideration the interests of diverse (target) constituencies for the business organization. | To serve as a means for obligatory taking into consideration the interests of diverse (target) constituencies for the business organization. |
| To serve as a means of increasing the company's business focus by officially providing a clear answer from the senior management in relation with pursued interests of certain constituencies and those that are rejected since an organization cannot be held responsible for everything. | To serve as a means of increasing the company's business focus by officially providing a clear answer from the senior management in relation with pursued interests of certain constituencies and those that are rejected since an organization cannot be held responsible for everything. |
| To outline basic directions for distribution of company resources. | |
| To ensure the realization of efficient communication with diverse constituencies for the company in order to guarantee the receiving of all needed resources for maintenance of its activities. | To ensure the realization of efficient communication with diverse constituencies for the company in order to guarantee the receiving of all needed resources for maintenance of its activities. |

Source: (Bartkus, Glassman, McAfee, 2000; Bart, Bontis, Taggar, 2001; Bart, 1997; Sanchez, Heene, 2004; Desmidt, Princie, Decramer, 2011).

● Undisciplined and interchangeable use of many constructs close to "mission" is detected in the scientific literature (Dermol, Breznik, 2012; Strong, 1997) and in many cases these constructs are not accompanied by respective definitions. The aggregate of these terms includes credo statement, purpose declaration, basic principles, guiding company activities, declaration of corporate intentions or firm vision, core values, philosophy, reason of business being, a means of image creation, a factor of differentiation, etc. In this way the joint confession by managers, consultants and scientists that there is no concord in relation with the shared meaning of the term mission, comes to the foreground.

● Highly appreciating the role of the mission statement in the strategic planning process, realized in the company which contributes to: (a) associating the mission with formulation of organizational goals; (b) perceiving the mission as an opportunity to search for specific answers for each entity to questions as: "What is the business of this organization?" and "What should be the business of this organization?"; (c) exploring the relationship among mission components (or characteristics) and organizational performance as a whole (Babnik, Breznik, Dermol, Širca, 2014).

● Making mission dependent on dominating cultural characteristics in the firm. In this way it accumulates nuances in its meaning as a declaration of organizational philosophy, identity and values, embedded in the pursued aims, kept norms, decisions made, undertaken actions and demonstrated everyday employee behaviors (Babnik, Breznik, Dermol, Širca, 2014).

● Adopting a perfunctory and succinct disclosure of mission that leads to its description as: (a) an instrument for achievement of internal company objectives (strategy design, performance appraisal, etc.) (Drucker, 1977; Klemm, Sanderson, Luffman, 1991); (b) an instrument for pursuing external communication objectives in order to model the perceptions of target constituencies (Campbell, 1997; Bartkus, Glassman, McAfee, 2000); (c) the first step in strategic planning (David, 1989); (d) a prerequisite for doing business (Smith, Heady, Carson, Carson, 2001); (e) an official document that is used for expressing a long-term and distinctive aim for a business organization (Rigby, Bilodeau, 2009).

● Literature reviews, thematic classification and summarizing of scientific publications permit identifying important nuances in the meaning of the term "mission" (Rajasekar, 2013): (a) orientation to constituencies; (b) orientation to identity clarification of clients, geographical markets, produced products and delivered services, the utility of the company, and the use of implemented technologies; (c) orientation to declaring of company philosophy (beliefs, values, desires, ethical priorities); (d) orientation to expressing company's comprehension in relation to possessed distinctive competence and competitive advantage; (e) orientation to design of specific characteristics for the mission as scope, level of abstractness, text length, etc.

Vision seems to be the second most popular company document for description of professed culture characteristics according to literature reviews (Davidson, 2002; Lipton, 2003; Lipton, 1996). It is defined as a specific impact that dominating ideology and management philosophy among decision-makers exercise on the respective business organization (Lynch, 2000). In this way the vision becomes a prerequisite for ensuring high quality design of future posible and desireable state of things or formulation of long-term aim for the respective organization. Thus, vision may be viewed as a kind of catalogue for organizational purpose and its overall development strategy. Hussey (1998) adds new aspects to vision's meaning, i.e. the set of professed company values. As far as the relationship „mission - vision" is concerned still there is no consensus about which of the two constructs is wider in scope or primary in comparison to the other one, and has a higher level of abstractness or on what is based the semantic relationship between them and even whether such relationship exists (see Hussey, 1998; Campbell, Tawadey, 1992; Sheaffer, Landau, Drori, 2008; Campbell, Yeung, 1991). One of the best models of illustrating the components of company vision and thus operationalizing it to satisfy management needs of ensuring a unique strategic direction for the business organization is shown below (see table 6).

**Table 6.** *Mark Lipton's model of company vision*

| Themes of vision | Finding unique answers to specific questions within the theme... |
|---|---|
| Mission | What business(es) are we in?<br>What is our fundamental purpose or reason of being?<br>What types of products or services do we make or provide?<br>How do we define the customers we serve?<br>For whose benefit are all our efforts?<br>What unique value do we bring to our customers?<br>Are we confident that this mission is distinct and unique from any other organization that may provide a similar product or service?<br>Are we describing what we do or why we do it? |
| Strategy | What is the basic approach to achieving the mission?<br>What is the distinct competence or competitive advantage that will characterize our organizational or departmental success? |
| Culture | What are (or should be) the hallmarks of our culture and leadership style?<br>How do (or should) we treat each other and how should we work together?<br>What do we believe about ourselves?<br>What do we stand for?<br>What values do we hold dear?<br>What characterizes the effective employee?<br>In what ways is our organization a great place to work? |

*Source: (Lipton, 2003; Lipton, 1996).*

## 3. Aspects of development of Bulgarian large business and local holding companies during the transition period (1990-2017)

The development of the national economy during the period 1990-1998 is characterized by initiation of the transition from the "planned centralized" to "market" economy with democratic political system where there is free competition between businesses and the prices of goods and services are determined by supply and demand on the market. This transition has been severely hampered by the legacy negatives in the economic, business cultural and political life of Bulgaria which predetermined the serious lagging behind of Bulgarian economy in comparison with many other countries from the Eastern European region in the processes of privatization and restitution of property, restructuring and renewal of local industry, NATO and EU accession and further integration in these supranational institutions. In this realm the first private holding companies came into being and demonstrated active participation in the mass privatization process (1996), the forced privatization government campaign (1998-2001) when the ownership of the Bulgarian industry was transformed from state-owned to private, and the subsequent process of redistribution of strategic economic entities among these business organizations (2001-2004), undertaken for the purpose of portfolio optimization. The impact of the World financial and economic crisis on Bulgarian economy and local holding companies (from 2008) may be described as major decreases in contracted orders by foreign clients, greater restraints on the provision of external financing for the business organization by the financial sector and delayed payments to private entities by the state in relation with public tenders. Because of this negative impact the managers of large holding companies were compelled to undertake certain organizational changes in order to survive in the short run and prosper in the long run (for example downsizing, closing some strategic economic entities, introducing innovation, changing applied business models, etc.).

In general the 1990s in Bulgaria are marked by sharp and frequent political crises and deep local economic depression, because politicians, opinion leaders, managers and scientists could not reach a consensus on the appropriate aims of development for the local industrial sector that deprived them from establishing a healthy discipline in thinking and undertaking consecutive actions at different levels (see Dochev, 1996; Tsvetkov, 1996; Georgiev, 1996; Kaligorov, 1996). However, a much more significant development in the Bulgarian industry could be observed during the first decade of the 21st century and until the present moment, despite the negative impact of the world financial and economic risis that is evident from data, describing turnover in the industrial sector, value of production, manufactured by the industrial sector, gross added value and gross domestic product in the industrial sector (see figure 3÷6). On the one hand, the observed high concentration of ownership in local holding companies reduced the positive effects of capital markets and established control systems on management discretionary behaviors. On the other hand, most of the banks in the country are owned by foreign investors who are to some extent cautious in their collaboration with domestic business entities, affiliated to specific business groups (Mueller, Dietl, Peev, 2003). Furthermore, large companies in Bulgaria were affected by increasing discrepancy between talent supply and demand on local labor market. That is why the lack of highly skilled specialists appears to be one of the main problems that local holding companies faced during the time period between 2000-2011 in congruence with insufficient orientation to employment in high value adding industries (Mitra, Pouvelle, 2012).

**Figure. 3.** *Turnover in the industrial sector for the period 2005-2015*

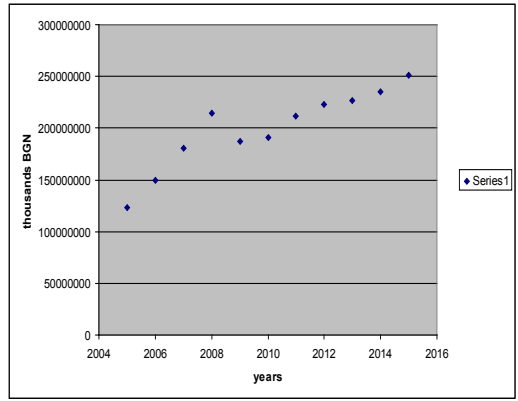

*Source: (\*\*\*, 2017h; \*\*\*, 2014).*

**Figure 4.** *Value of production, manufactured by industrial sector for the period 2005-2015*

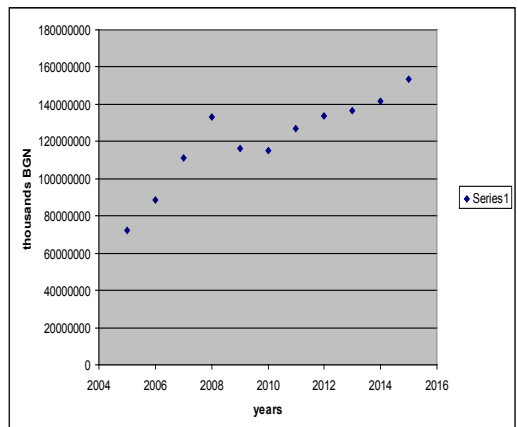

*Source: (\*\*\*, 2017h; \*\*\*, 2014).*

**Figure 5.** *Gross added value in the industry sector for the period 2000-2015*

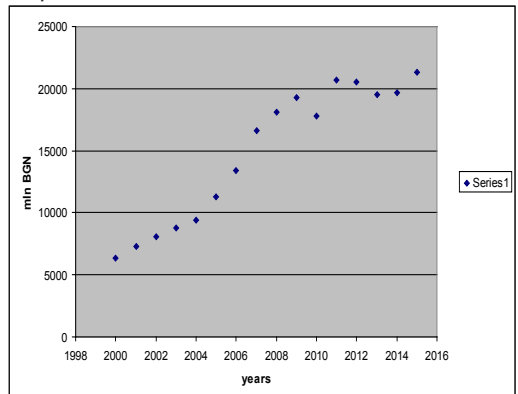

*Source: (\*\*\*, 2017h; \*\*\*, 2014).*

**Figure 6.** *GDP of the industrial sector for the period 2000-2015*

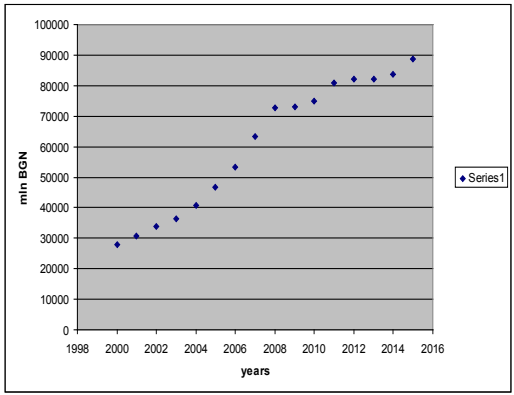

*Source: (\*\*\*, 2017h; \*\*\*, 2014).*

That is why it may be considered that our country had no real opportunities for strengthening domestic consumption of industrial products during the analyzed period. Low households income on average is another stopping factor, counteracting potential increases in the consumption of industrial products. Therefore export companies are the only possible source of industrial growth. For this reason the export orientation of Bulgarian firms is extremely important, bringing to the playground specific challenges, as follows (Iliev, 2013; Dimitrov, 2012b):

● High specialization of production processes, manufactured products and delivered services.

● A narrow and unfavorable structure of industrial exports.

● Concentrated orientation of exports to some European countries.

● Insufficient economic impact.

● High level of intercompany indebtedness.

● Significant decline in the quality of human resources at company level.

● Complicated and broken relations and interaction among companies.

● The necessity of entering and saturating the entire single European market by big local holding companies, starting the intensive relations with partners from European regions closer in dominating cultural attitudes to business (see Hofstede, Hofstede, Minkov, 2010; Trompenaars, Hampden-Turner, 1998; Paunov, 2009; Dimitrov, 2012b).

● The necessity of entering and integrating in sub-contractor chains and entrepreneurship networks with similar cultural attitude to doing business by local small and medium-sized business organizations (see Todofov, 2011).

● Low own investments and public spending in the sphere of research and innovations.

● Incurring the highest energy costs and the lowest efficiency within the EU.

The prevailing resource consumption character is the main feature of the Bulgarian industry which acts against industrial growth because in recent years the resources, both tangible and intangible, have depleted and become more expensive. Leading industrial countries reoriented their production to less energy intensive and highly innovative technologies. Unfortunately transition in Bulgarian industry from energy consuming to more energy-efficient production is slow and complicated. The prolonged economic and financial crisis may be viewed as an additional difficulty to the implementation of such transition by the local holding companies. It is evident that accelerated industrial growth in Bulgaria cannot occur in the near future, but it is necessary to design a strategy for industrial growth, based on a long-term vision for the development of the Bulgarian economy. These deliverables should be based on thorough analysis, concerning the main problems for local holding companies in such transition, as not so well developed management concepts for cost optimization in our business organizations, insufficient use of different bonus schemes to motivate staff and serious decline in investments (Iliev, 2013).

The situation for export-oriented firms is further aggravated by the highly saturated markets with domestic and foreign products. Studies reveal that export companies overcome the negative effects caused by the economic crisis to some extent, but since those companies implement humble investment programs tangible results in improving competitiveness are not achieved (Lin, Chang 2009; Shafaeddin, 1998).

Recent years, however, confirmed the inability to rely solely on market forces and competition to design and implement the needed restructuring of local industry. The arbitrary withdrawal executed by the state administration from many sectors of the Bulgarian industry has been one of the main reasons for the lack of industrial growth during the transition period. That is why the joint efforts of business owners and management of the companies, represented by employer organizations, representatives of the financial, banking and other non-governmental organizations as well as opinion leaders from the academic community, closely related to the industrial sector, are needed in order to formulate and implement a set of adequate measures (Iliev, 2012).

The necessity of strategic vision for development of Bulgarian industry brings forth some controversial issues. To what extent does the Bulgarian government possess a sufficient potential to successfully support the Bulgarian economy? Unfortunately, data analysis from recent years shows that the state administration has no sufficient capacity to be an equal and adequate business partner. A significant change in the interaction between state administration and the business entities is urgently needed (Iliev, 2013). This can be achieved by determining the directions where realistic strategic decisions for industrial growth can be sought with the active support of the state (for example, the industry's product orientation). The penetration into international markets must be based on complex research, aimed at searching for appropriate commodities, conquering high levels of competitiveness for local industrial enterprises on foreign markets. The product orientation can be the reason for support on the part of the government institutions, and direct their commitment to industrial growth, adhering to EU rules and regulations. Such strategic orientation does not imply privileged positioning of separate sub-sectors and industries. Several main arguments in support of the equal importance of certain subsectors and productions can be pointed out (Lall, 1994; Todorov, 2011; Iliev, 2013, 2008):

● Inability for companies with low investment and credit potential to attract foreign investors or financial institutions, as well as to realize their own large-scale investment projects.

● Companies with the potential to become part of international production chains need government support to be able to meet the high demands of consumers and become an integral part of those chains.

● Being a part of the external markets for resource consuming industries is very difficult. In order to successfully enter the foreign markets, it is necessary for the branches of the export industry to have optimized costs, as well as increased investment activity in the sector.

● Smart orientation in growth stimulation to both target groups – the big holding companies, and small and medium-sized enterprises, is needed.

With regard to strategic product orientation for local firms considerable attention must be paid to numerous important aspects such as (Iliev, 2013; Weiss, 2011; Dimitrov, 2011c):

● Making use of the opportunities, set out in EU industrial policy (see ***, 2016).

● Stepping up innovation and investment activity by the state in order to promote and support industrial growth.

- Increasing trust between banks and business by setting up channels for interaction between the financial sector and the manufacturing sector.

- Establishment of a guarantee fund to support small and medium-sized enterprises, including in their potential collaboration initiatives with big holding companies.

- Introducing a consistent development of the local industry with a leading product focus and high added value generation.

With regard to the state institutions and their participation in the processes of economic growth, it can be said that the state should support and not intervene in the generation and regulation of the economic growth in the Bulgarian economy. State support should involve attracting foreign investors and investments, supporting the creation of an infrastructure for applied research and innovation, and especially in reforming the education system. The main developmental trends in the education system should be to renew, and in some cases, to establish vocational high schools, specialized in training of qualified staff for the contemporary industrial sectors, to generate the highest value added, as well as to create legislative mechanisms (including financial levers) to strengthen the relationship "university – business organizations" (Iliev, 2008).

The World financial and economic crisis marks a key moment in the development of the Bulgarian industry (2008-2013), because this is the first deep crisis after the transition to a market economy for the countries from Eastern Europe, bearing features that directly or indirectly negatively affected Bulgaria (Dimitrov, 2011b; Iliev et.al 2011):

- The global span and duration of the current crisis increased the risks for the Bulgarian economy.

- Due to the fact that the economic crisis is based on major economic imbalances, it has caused a collapse of liquidity and insolvency for many businesses all over the world, which creates great uncertainty.

- Devaluation of currencies in a number of countries where local organizations have suppliers, partners and clients.

For Bulgaria, the real risk of adverse effects from the global crisis has arisen not only because of a slowdown and a significant reduction in the inflow of foreign investment, but also due to the continued contraction of external markets for local exporting firms, consequently observed decreases in investment activities and personnel employment, and banks' more restrictive lending policy. The crisis hit Bulgaria a little bit later in comparison to the other EU countries, providing an opportunity to formulate and coordinate a policy for overcoming the crisis, bearing in mind the tranquility, resulting from the implementation of the currency board. This policy could go beyond short-term anti-crisis measures, focusing on the problems accumulated in the economy, and encompass the full array of contemporary anti-crisis management. This crisis should be used as an opportunity of restructuring the national economy in order to counterbalance the accumulated large imbalances in the long run and should be considered and reflected in applied professed corporate culture attributes (Iliev et.al, 2011, Dimitrov, 2012a; Dimitrov, 2011a; Dimitrov, 2007). Manufacturing, trading, mining and construction are the most affected industries by the economic crisis. The behavior of companies depends to a great extent on their assessment of the duration of the crisis. If companies expect the crisis to be shorter, they will take measures to provide them with the necessary liquidity to meet current liabilities without undertaking long-term business restructuring activities and performing serious changes in their professed culture attributes. If, according to the firms, the crisis persists for a longer time, then they undertake structural changes to increase the efficiency of production, mainly related to the rational use of resources and the production capacity, and deliberately initiate changes in the official company culture. The decrease in resource prices increases the freedom of companies to manage their resources. So the managers can choose between two options – the first one is to reduce the prices of the final output, i.e. to improve their price competitiveness and to try to maintain or expand their market share, while the other option is to keep the prices of the final products and have more financial resources that depend on their dominating cultural characteristics (Iliev et.al, 2011, Dimitrov, 2008). Already presented statistical data (see figure 3÷6) reveals that senior managers in many local business organizations dealt appropriately with the effects from the crisis, but the generated growth was still not enough to correspond to the market performance levels of the leading companies from the elder member states of the EU, so it seems that the crisis was not efficiently used as a key marker event for realizing appropriate organizational changes in order to solve pending business issues, including cultural ones.

## 4. Survey methodology

The aim of the survey is to make a snapshot of the peculiarities and detected nuances in structuring and expressing of the professed culture in the virtual realm among the group of holding companies, members of the Bulgarian Industrial Capital Association[2]. The lack of exisitng surveys in this sphere in Bulgaria predetermines the choice of the researchers not to apply and explore the availability of ideal professed firm culture scheme among companies of any kind, elaborated by renowned foreign and Bulgarian scientists. In this way even the minimal or partial management efforts of organizational decision-makers may be detected and considered through the performed analysis, i.e. even fragments of official company documents. That is how the list of official company documents, intended for description of professed corporate culture, was created (see the beginning of section 2 in this article) and their potential and specific presence on the respective websites was carefully appraised, relying on the achievement of prominent scientists in this field (see Van Nimwegen, Bollen, Hassink, Thijssens, 2008; Leuthesser, Kohi, 1997; Campbell, Shrives, Bohmbach-Saager, 2001; Darbi, 2012). Furthermore, the aim of the survey was decomposed to several tasks, as follows:

● Identification of concrete types of official company documents, used to describe existing professed culture characteristics.

● Identification of the target constituencies, associated with the presented content of the aforementioned official company documents.

● Identification of dominating language versions of the company websites, applied for intentional cultural disclosure of respective entities.

● Identification of professed company values and management principles on these sites, associated with key decision-making by senior management.

● Performing an analysis of the two most popular official company documents – mission and vision, applied for description of professed culture from the firms in the virtual realm. The focus here is set on searching for their structure, semantic scope, length in number of used words.

● Collection of general data for the surveyed holding organizations (i.e. one-dimensional distributions and cross tabulations of selected variables).

Limitations of the survey were also formulated. Some of these facilitated the work of the researchers, while others were imposed by the business environment, and are as follows:

---

[2] This survey is a part of a larger scientific project, aiming at identification of dominating professed corporate culture characteristics among business entities, operating in Bulgaria (see Dimitrov, Ivanov, Geshkov, 2016, 2015).

- Financial data for the holding companies, encompassing at least three consecutive annual periods (profit, net turnover) could not be found for the majority of holding companies, although it should be officially declared on the site of the Registration agency (***, 2017i) and it was sometimes available on the site of the respective company. That is why it was not possible to relate the possession of professed corporate culture attributes with the financial performance of these big business organizations.

- The companies whose websites were not accessible for viewing, were excluded from the analysis.

## 5. Results from the empirical survey of professed culture attributes, possessed by the members of the Bulgarian Industrial Capital Association

With regard to the identified official business documents, the content of which is found on the relevant websites of the business organizations surveyed, it is clear that the provision of generic and concise information, describing produced products and/ or delivered services by the respective companies, is the top priority for the majority of their senior managers. This situation is typical of the holding companies, belonging to the Bulgarian Industrial Capital Association. That is why the most popular of the documents, examined among the companies revealing at least one official document in the sphere of professed culture, is „about us..." (90.9%). „Vision" (13.6%), „mission" (9.1%) and „corporate/ our values" (9.1%) are relatively less common among these business organizations. Professed corporate culture official documents as „motto", „credo", „corporate/ corporate/ official philosophy/ policy", „our (organizational) history", „ethical code", „corporate organizational culture","corporate social responsibility", "sustainable company level development", „slogan" and „manifesto" are not mentioned within this group (see table 7 and figure 7).

Table 7. *Most used official professed culture documents on the websites of the surveyed companies*

**$x_1m$ Frequencies**

|  |  | Responses | | Percent of Cases |
|---|---|---|---|---|
|  |  | N | Percent |  |
| Professed culture - documents[a] | Vision | 3 | 11,1% | 13,6% |
|  | Mission | 2 | 7,4% | 9,1% |
|  | Firm/ our values | 2 | 7,4% | 9,1% |
|  | About us... | 20 | 74,1% | 90,9% |
| Total |  | 27 | 100,0% | 122,7% |

a. Dichotomy group tabulated at value 1.

**Figure 7.** *Most popular professed culture company documents, presented on the websites of the surveyed entities*

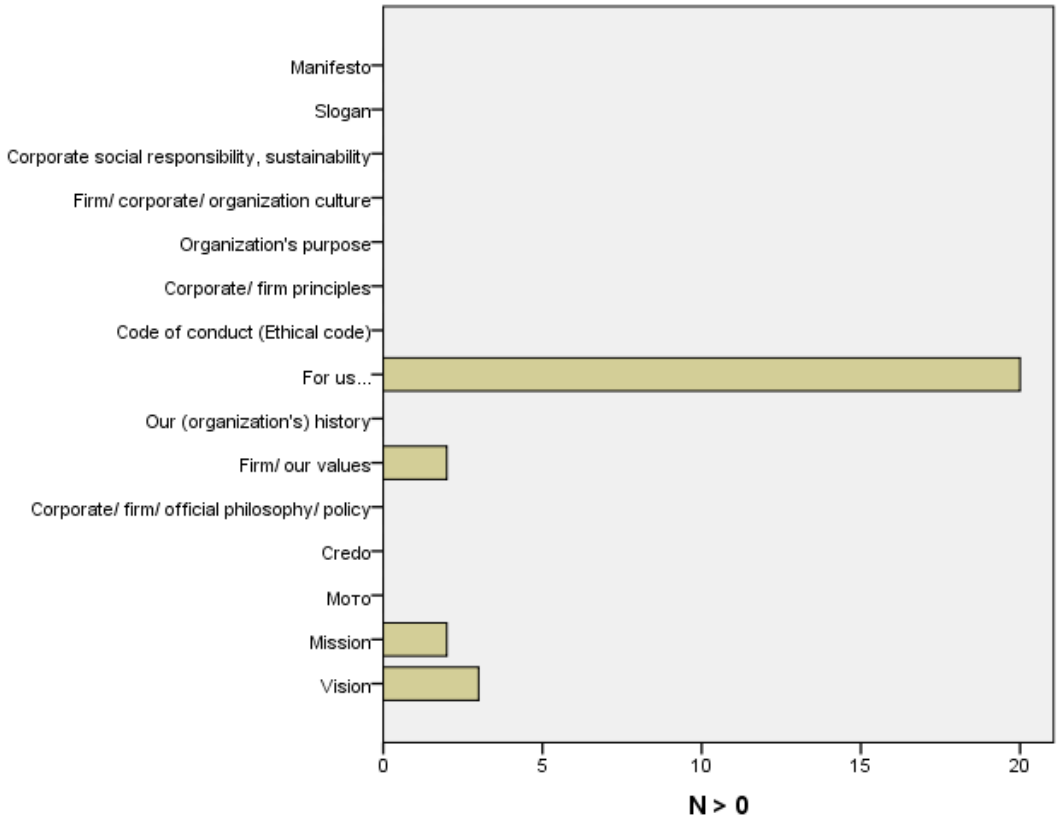

With regard to stakeholders that are mentioned in the contents of the official company documents, presented on the respective company websites, the most frequently mentioned ones are local community (100%), managers (90.9%), employees (77.3%), suppliers (77.3%), clients, customers (68.2%). Stakeholders as Investors & shareholders/ stockholders (45.5%), Competitors (40.9%), Creditors (27.3%), Ecologic movements (22.7%) and Board members (22.7%) are characterized by lower popularity within this group of entities. Organized labor, trade unions and other NGOs are not mentioned (see table 8 and figure 8).

**Table 8.** *Stakeholders cited on the companies' websites*

**$x_2m Frequencies**

| | | Responses | | Percent of Cases |
|---|---|---|---|---|
| | | N | Percent | |
| Constituencies[a] | Media & opinion leaders | 2 | 1,6% | 9,1% |
| | Local community | 22 | 17,1% | 100,0% |
| | Government/ Regulators | 1 | 0,8% | 4,5% |
| | Board members | 5 | 3,9% | 22,7% |
| | Suppliers | 17 | 13,2% | 77,3% |
| | Clients, customers | 15 | 11,6% | 68,2% |
| | Investors & shareholders/ stockholders | 10 | 7,8% | 45,5% |
| | Ecologic movements | 5 | 3,9% | 22,7% |
| | Competitors | 9 | 7,0% | 40,9% |
| | Managers | 20 | 15,5% | 90,9% |
| | Employees | 17 | 13,2% | 77,3% |
| | Creditors | 6 | 4,7% | 27,3% |
| Total | | 129 | 100,0% | 586,4% |

a. Dichotomy group tabulated at value 1.

**Figure 8.** *The most popular stakeholder groups, mentioned in the target professed company culture documents, applied by the surveyed business organizations on the internet*

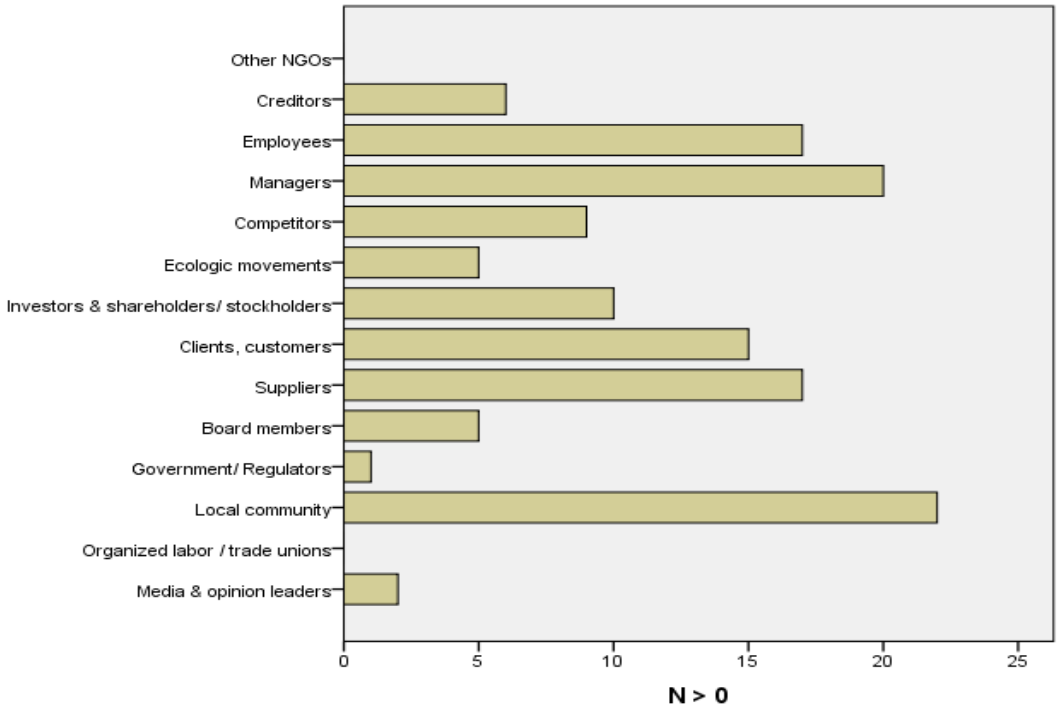

The survey results show that Bulgarian language (100%), followed by English language (90.9%), are the most widely used languages on the holding companies' websites. There are no significant differences in the website content between the Bulgarian and the English versions for the majority of the surveyed companies (77.3%) (see table 9 and table 10). Just two local holding companies give priority to the professed culture information, provided in Bulgarian language. The majority of the holding companies maintain both language vesions – Bulgarian and English, for their websites.

**Table 9.** *Languages used on the websites of the companies*

**$x\_3m$ Frequencies**

| | | Responses | | Percent of Cases |
|---|---|---|---|---|
| | | N | Percent | |
| Language versions[a] | Bulgarian language | 22 | 52,4% | 100,0% |
| | English language | 20 | 47,6% | 90,9% |
| Total | | 42 | 100,0% | 190,9% |

a. Dichotomy group tabulated at value 1.

**Table 10.** *Differences in Bulgarian and English version of the companies' websites*

**Are any differences in the presented information, concerning professed culture in Bulgarian and English language detected?**

| | | Frequency | Percent | Valid Percent | Cumulative Percent |
|---|---|---|---|---|---|
| Valid | no differences | 17 | 77,3 | 77,3 | 77,3 |
| | the content in Bulgarian language is richer | 2 | 9,1 | 9,1 | 86,4 |
| | not applicable | 3 | 13,6 | 13,6 | 100,0 |
| | Total | 22 | 100,0 | 100,0 | |

The major official company values, pointed out on the respective websites, are: flexibility (100%), excellence (100%) and innovativeness (innovation) (100%) while other values such as efficiency, inspiration, honesty, dynamics, responsibility, commitment to customers, talent, integrity, (empowerment &) accountability, financial responsibility, collaboration, velocity and value are mentioned just once (see table 11). It seems that diversity among the company members of this group is great, concerning the observed officially declared values. On the one hand, this situation may create a potential for realization of cultural misunderstandings that may harm collaboration among the holding companies when formulating and implementing strategies at network level. On the other hand, these widespread values among holding companies may be partially explained by their operating in different industries (by detailed classification).

**Table 11.** *Professed company values on the websites of the holding companies*

**$x\_7m Frequencies**

| | | Responses | | Percent of Cases |
|---|---|---|---|---|
| | | N | Percent | |
| Professed firm values[a] | Efficiency | 1 | 5,3% | 50,0% |
| | Inspiration | 1 | 5,3% | 50,0% |
| | Innovativeness/ innovation | 2 | 10,5% | 100,0% |
| | Honesty | 1 | 5,3% | 50,0% |
| | Dynamics | 1 | 5,3% | 50,0% |
| | Responsibility | 1 | 5,3% | 50,0% |
| | Commitment to customers | 1 | 5,3% | 50,0% |
| | Flexibility | 2 | 10,5% | 100,0% |
| | Excellence | 2 | 10,5% | 100,0% |
| | Talent | 1 | 5,3% | 50,0% |
| | Integrity | 1 | 5,3% | 50,0% |
| | (Empowerment &) Accountability | 1 | 5,3% | 50,0% |
| | Financial responsibility | 1 | 5,3% | 50,0% |
| | Collaboration | 1 | 5,3% | 50,0% |
| | Velocity | 1 | 5,3% | 50,0% |
| | Value | 1 | 5,3% | 50,0% |
| Total | | 19 | 100,0% | 950,0% |

a. Dichotomy group tabulated at value 1.

The proclaimed management principles are not used as a method to describe a target firm culture on the internet by the local holding comanies. Furthermore, only 9.1 % of the companies use the "mission" and "vision" documents simultaneously on their websites for the purpose of describing their specific firm culture. The significant difference between the applied terms "mission" and "vision" is clearly outlined in 4.5% of the surveyed companies. Among another 4.5% of the companies a significant difference between the applied terms "mission" and "vision" cannot be outlined. The majority of holding companies (20 in number) do not apply mission and vision simultaneously and that is why "not applicable" option is chosen for them (see table 12).

**Table 12.** *The difference between the terms "mission" and "vision"*

**Is the difference between the applied terms "mission" and "vision" clearly outlined?**

| | | Frequency | Percent | Valid Percent | Cumulative Percent |
|---|---|---|---|---|---|
| Valid | Yes | 1 | 4,5 | 4,5 | 4,5 |
| | No | 1 | 4,5 | 4,5 | 9,1 |
| | Not applicable | 20 | 90,9 | 90,9 | 100,0 |
| | Total | 22 | 100,0 | 100,0 | |

From the site of one holding company it can be inferred that the "mission" contains the essence of the vision. For the rest 95.5% of the companies such a conclusion cannot be drawn. Among the components of the company mission that may be identified in the present content the following could be distinguished: specification of the target markets (94,7%) and principal products and/ or services (94,7%), followed by geographic domain (89.5%) and competitive strategy (89.5%). Behavioral standards are the third most popular among the components of the company mission with 78.9% (see table 13 and figure 9). Mission components as commitment to survival, growth and profitability, key elements in company philosophy, desired public image, purpose/ goal/ aim of the organization and organization results are not applied here. The rest of the mission components are used sporadically.

**Table 13.** *Components of mission statements, applied by surveyed companies*

**$x\_12m$ Frequencies**

| | | Responses | | Percent of Cases |
| --- | --- | --- | --- | --- |
| | | N | Percent | |
| Components of company mission[a] | Specification of target markets | 18 | 18,2% | 94,7% |
| | Principal products and/ or services | 18 | 18,2% | 94,7% |
| | Geographic domain | 17 | 17,2% | 89,5% |
| | Core technology | 2 | 2,0% | 10,5% |
| | Company self-concept, identity | 1 | 1,0% | 5,3% |
| | Desired competitive position | 3 | 3,0% | 15,8% |
| | Competitive strategy | 17 | 17,2% | 89,5% |
| | Behavioral standards | 15 | 15,2% | 78,9% |
| | Reason for being | 1 | 1,0% | 5,3% |
| | Distinctive competencies | 1 | 1,0% | 5,3% |
| | Clear competitive advantages | 1 | 1,0% | 5,3% |
| | Clearly defined interests of the organization | 2 | 2,0% | 10,5% |
| | Specific financial objectives, targets | 1 | 1,0% | 5,3% |
| | Specific non-financial objectives, targets | 1 | 1,0% | 5,3% |
| | Core ideology | 1 | 1,0% | 5,3% |
| Total | | 99 | 100,0% | 521,1% |

a. Dichotomy group tabulated at value 1.

**Figure 9.** *Dominating components of mission statements in the group of investigated holding companies*

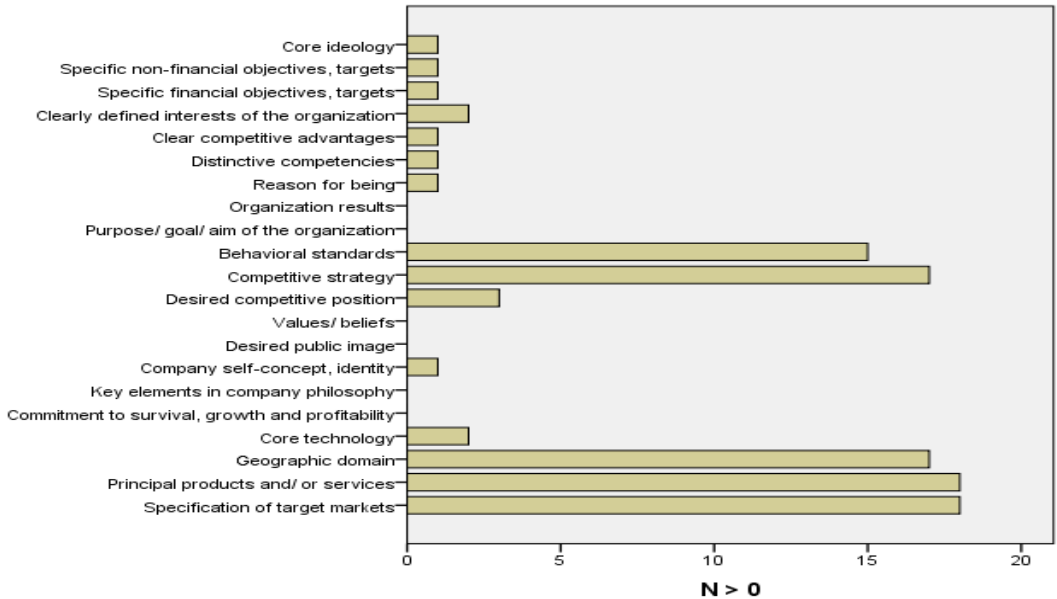

The major components of a company's vision, that could be outlined, are: company's strategy (44.4%) and envisioned future (44.4%), followed by core ideology (5.6%), corporate culture (2.8%) and company mission (2.8%). With regard to the observed length of the company's mission it can be concluded that it has just two values - 73 words and 33 words. The situation with the vision's length is similar - two cases with 16 words and 24 words. Furthermore, about 31.8% of the surveyed companies maintain diverse, specific and different key attributes of their professed culture on the (sub-)sites of the headquarters and subsidiaries. The other 68.2% of the holding organizations seem not to maintain such differences in electronic content on company (sub-)sites or do not have websites for their subsidiaries.

In this survey about 72.7% of the business organizations contribute to the services sector, about 22.7% of them operate in the industry sector and about 4.5% of them function in the financial sector (see table 14).

**Table 14.** *Distribution of companies to respective industries*

**What sector of the economy does the business organization contribute to?**

| | | Frequency | Percent | Valid Percent | Cumulative Percent |
|---|---|---|---|---|---|
| Valid | Industrial production | 5 | 22,7 | 22,7 | 22,7 |
| | Services | 16 | 72,7 | 72,7 | 95,5 |
| | Financial services | 1 | 4,5 | 4,5 | 100,0 |
| | Total | 22 | 100,0 | 100,0 | |

The two-dimensional distributions reveal even richer nuances in the general picture of the indicated culture characteristics, emerging among the surveyed companies. For example, the holding companies within the industrial production sector apply only the official document "about us…", while the situation in the services sector is much richer in terms of the observed use of different types of appropriate official documents – "About us..." (87.5%), "Vision" (18.8%), "Mission" (12.5%) and "Firm's/ our values" (12.5%). Just one company belongs to the financial services sector and it applies the simple option of "about us…" (table 15 and figure 10).

**Table 15.** *Cross-tabulation between professed culture documents and industrial sector of surveyed holding companies*

**$x_1m*x_24 Crosstabulation**

| | | | What sector of the economy does the business organization contribute to? | | | Total |
|---|---|---|---|---|---|---|
| | | | Industrial production | Services | Financial services | |
| Professed culture - documents[a] | Vision | Count | 0 | 3 | 0 | 3 |
| | | % within $x_1m | 0,0% | 100,0% | 0,0% | |
| | | % within x_24 | 0,0% | 18,8% | 0,0% | |
| | | % of Total | 0,0% | 13,6% | 0,0% | 13,6% |
| | Mission | Count | 0 | 2 | 0 | 2 |
| | | % within $x_1m | 0,0% | 100,0% | 0,0% | |
| | | % within x_24 | 0,0% | 12,5% | 0,0% | |
| | | % of Total | 0,0% | 9,1% | 0,0% | 9,1% |
| | Firm's/ our values | Count | 0 | 2 | 0 | 2 |
| | | % within $x_1m | 0,0% | 100,0% | 0,0% | |
| | | % within x_24 | 0,0% | 12,5% | 0,0% | |
| | | % of Total | 0,0% | 9,1% | 0,0% | 9,1% |
| | About us... | Count | 5 | 14 | 1 | 20 |
| | | % within $x_1m | 25,0% | 70,0% | 5,0% | |
| | | % within x_24 | 100,0% | 87,5% | 100,0% | |
| | | % of Total | 22,7% | 63,6% | 4,5% | 90,9% |
| Total | | Count | 5 | 16 | 1 | 22 |
| | | % of Total | 22,7% | 72,7% | 4,5% | 100,0% |

Percentages and totals are based on respondents.

a. Dichotomy group tabulated at value 1.

**Fig. 10.** *Distribution of identified professed culture documents by industrial sector for the surveyed holding companies*

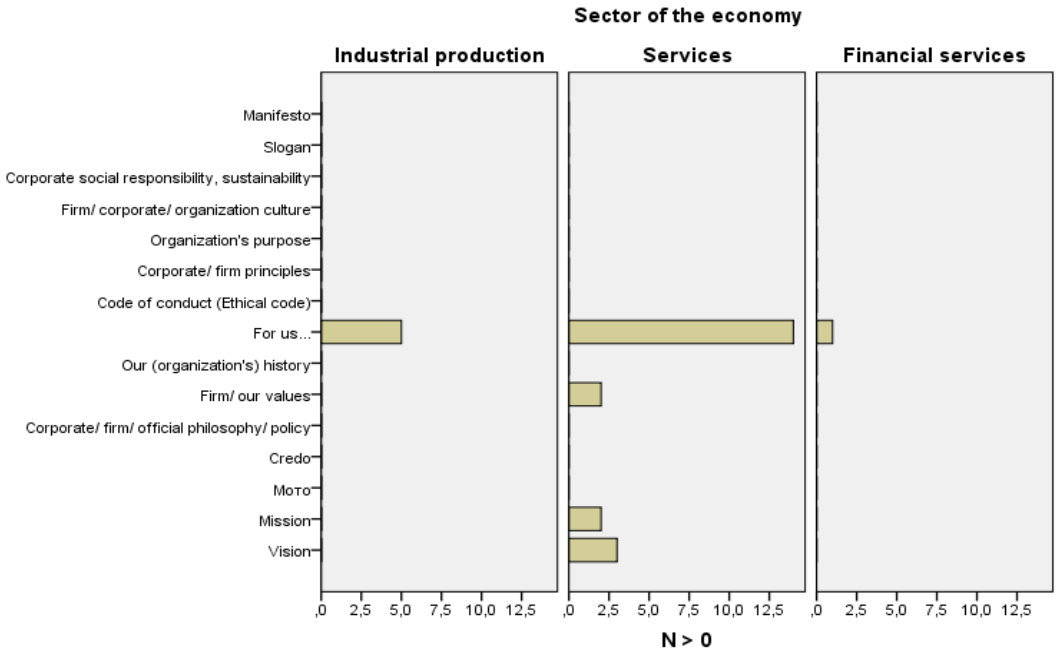

The most popular constituencies for the holding companies within the industrial production sector are „Local community" (100%), „Managers" (80%), „Suppliers" (60%), „Clients, customers" (60%), „Investors & shareholders/ stockholders" (60%) and „Employees" (60%). Here, constituencies as „Media & opinion leaders", „Government/ Regulators", „Creditors", „Organized labor/ trade unions" and „Other NGOs" are not mentioned. The other constituencies from the list are rarely applied by this sub-group of companies. The richest picture, regarding addressed constituencies, is observed within the services sector where the most popular ones are „Local community" (100%), „Managers" (93.8%), „Suppliers" (81.3%), „Employees" (81.3%) and „Clients, customers" (68.8%). Just two items from the preliminary generated list of constituencies are not used here – „Organized labor/ trade unions" and „Other NGOs". The financial sector, presented only by one holding company, reveals single use of several constituencies – "Local community", "Suppliers", "Clients, customers", "Managers" and "Employees" (table 16 and figure 11).

**Table 16.** *Cross tabulation between constituencies and industries to which companies contribute*

**$x\_2m*x\_24 Crosstabulation**

| | | | What sector of the economy does the business organization contribute to? | | | Total |
|---|---|---|---|---|---|---|
| | | | Industrial production | Services | Financial services | |
| Constituencies[a] | Media & opinion leaders | Count | 0 | 2 | 0 | 2 |
| | | % within $x\_2m | 0,0% | 100,0% | 0,0% | |
| | | % within x\_24 | 0,0% | 12,5% | 0,0% | |
| | | % of Total | 0,0% | 9,1% | 0,0% | 9,1% |
| | Local community | Count | 5 | 16 | 1 | 22 |
| | | % within $x\_2m | 22,7% | 72,7% | 4,5% | |
| | | % within x\_24 | 100,0% | 100,0% | 100,0% | |
| | | % of Total | 22,7% | 72,7% | 4,5% | 100,0% |
| | Government/ Regulators | Count | 0 | 1 | 0 | 1 |
| | | % within $x\_2m | 0,0% | 100,0% | 0,0% | |
| | | % within x\_24 | 0,0% | 6,3% | 0,0% | |
| | | % of Total | 0,0% | 4,5% | 0,0% | 4,5% |
| | Board members | Count | 1 | 4 | 0 | 5 |
| | | % within $x\_2m | 20,0% | 80,0% | 0,0% | |
| | | % within x\_24 | 20,0% | 25,0% | 0,0% | |
| | | % of Total | 4,5% | 18,2% | 0,0% | 22,7% |
| | Suppliers | Count | 3 | 13 | 1 | 17 |
| | | % within $x\_2m | 17,6% | 76,5% | 5,9% | |
| | | % within x\_24 | 60,0% | 81,3% | 100,0% | |
| | | % of Total | 13,6% | 59,1% | 4,5% | 77,3% |
| | Clients, customers | Count | 3 | 11 | 1 | 15 |
| | | % within $x\_2m | 20,0% | 73,3% | 6,7% | |
| | | % within x\_24 | 60,0% | 68,8% | 100,0% | |
| | | % of Total | 13,6% | 50,0% | 4,5% | 68,2% |
| | Investors & shareholders/ stockholders | Count | 3 | 7 | 0 | 10 |
| | | % within $x\_2m | 30,0% | 70,0% | 0,0% | |
| | | % within x\_24 | 60,0% | 43,8% | 0,0% | |
| | | % of Total | 13,6% | 31,8% | 0,0% | 45,5% |
| | Ecologic movements | Count | 1 | 4 | 0 | 5 |
| | | % within $x\_2m | 20,0% | 80,0% | 0,0% | |
| | | % within x\_24 | 20,0% | 25,0% | 0,0% | |
| | | % of Total | 4,5% | 18,2% | 0,0% | 22,7% |
| | Competitors | Count | 2 | 7 | 0 | 9 |
| | | % within $x\_2m | 22,2% | 77,8% | 0,0% | |
| | | % within x\_24 | 40,0% | 43,8% | 0,0% | |
| | | % of Total | 9,1% | 31,8% | 0,0% | 40,9% |
| | Managers | Count | 4 | 15 | 1 | 20 |
| | | % within $x\_2m | 20,0% | 75,0% | 5,0% | |
| | | % within x\_24 | 80,0% | 93,8% | 100,0% | |
| | | % of Total | 18,2% | 68,2% | 4,5% | 90,9% |
| | Employees | Count | 3 | 13 | 1 | 17 |
| | | % within $x\_2m | 17,6% | 76,5% | 5,9% | |
| | | % within x\_24 | 60,0% | 81,3% | 100,0% | |
| | | % of Total | 13,6% | 59,1% | 4,5% | 77,3% |
| | Creditors | Count | 0 | 6 | 0 | 6 |
| | | % within $x\_2m | 0,0% | 100,0% | 0,0% | |
| | | % within x\_24 | 0,0% | 37,5% | 0,0% | |
| | | % of Total | 0,0% | 27,3% | 0,0% | 27,3% |
| Total | | Count | 5 | 16 | 1 | 22 |
| | | % of Total | 22,7% | 72,7% | 4,5% | 100,0% |

Percentages and totals are based on respondents.
a. Dichotomy group tabulated at value 1.

**Figure 11.** *Distribution of company constituencies by sectors of economy for the surveyed entities*

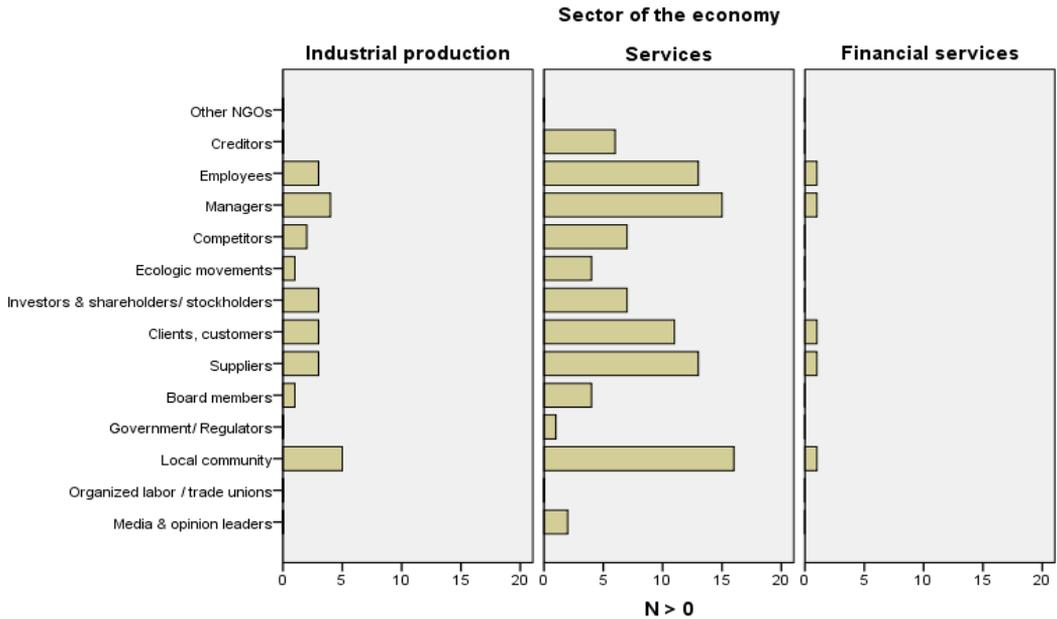

Bulgarian language is the most common among the websites of the companies operating in services sector (100%), while the holding companies from the other two sectors demonstrate a parity in using both the Bulgarian and English versions for their websites (table 17). It seems that just two holding companies have only Bulgarian versions of their websites. It is evident that 77.3% of the surveyed companies do not tolerate any differences in the presented information, concerning professed culture in Bulgarian and English.

**Table 17.** *Cross-tabulation between applied language version on the company site and sectors of industry where surveyed organizations operate*

**$x_3m*x_24 Crosstabulation**

| | | | What sector of the economy does the business organization contribute to? | | | Total |
|---|---|---|---|---|---|---|
| | | | Industrial production | Services | Financial services | |
| Language versions | Bulgarian language | Count | 5 | 16 | 1 | 22 |
| | | % within $x_3m | 22,7% | 72,7% | 4,5% | |
| | | % within x_24 | 100,0% | 100,0% | 100,0% | |
| | | % of Total | 22,7% | 72,7% | 4,5% | 100,0% |
| | English language | Count | 5 | 14 | 1 | 20 |
| | | % within $x_3m | 25,0% | 70,0% | 5,0% | |
| | | % within x_24 | 100,0% | 87,5% | 100,0% | |
| | | % of Total | 22,7% | 63,6% | 4,5% | 90,9% |
| Total | | Count | 5 | 16 | 1 | 22 |
| | | % of Total | 22,7% | 72,7% | 4,5% | 100,0% |

Percentages and totals are based on respondents.
a. Dichotomy group tabulated at value 1.

Just two holding companies from the services sector apply the values approach to outline key aspects of their specific corporate culture. The only shared values between these firms are „Innovativeness/ innovation", „Flexibility' and „Excellence". Other preofessed culture values are mentioned once by just one of these organizations, as follows: „Efficiency", „Inspiration", „Honesty", „Dynamics", „Responsibility", „Commitment to customers", „Talent", „Integrity", „(Empowerment &) Accountability", „Financial responsibility", „Collaboration", „Velocity" and „Value". Furthermore, the holding companies do not apply official management principles as a way of disclosing their professed culture on the internet.

The observations of company websites reveal that just two entities use the two terms „mission" and „vision" in the presented information, describing their culture, And the content analysis shows that only in one case it may be concluded that clearly outlined differences between the two constructs exist. In this case the mission contains the essence of the vision in itself.

On the sites of a sub-group of 19 holding companies there may be identified separate components of mission, even if there are no offically declared mission statements by the respective teams of senior managers. The use of mission components is most popular in the services sector where the most shared components are „Specification of target markets" (92.3%), „Principal products and/ or services" (92.3%), „Geographic domain" (84.6%), „Competitive strategy" (84.6%) and „Behavioral standards" (69.2%). Other components such as „Core technology", „Company self-concept, identity", „Desired competitive position", „Reason for being", „Distinctive competencies", „Clear competitive advantages", „Clearly defined interests of the organization", „Specific financial objectives, targets", „Specific non-financial objectives, targets" and „Core ideology" are rarely applied. The situation within the Industrial production sector is characterized by a higher level of concentration of applied components from all of the observed five holding companies, as follows: „Specification of target markets" (100%), „Principal products and/ or services" (100%), „Geographic domain" (100%), „Competitive strategy" (100%) and „Behavioral standards" (100%). The entity from the financial services sector shows full coincidence of chosen mission components with these, applied by the holding companies, belonging to the industrial production sector.

The observed situation with the vision components reveals that:

• In the Industrial production sector the holding companies apply only „strategy" (100%) and „envisioned future" (100%).

• The players from the services sector not only reproduce the same situation, but also give evidence to rare use of the other components of company vision as „mission", „firm culture" and „core ideology".

• The only entity, representing the financial services sector, replicates the behavior of the holding companies from the industrial production sector.

## Conclusion

The analysis reveals that the formulation of professed cultural characteristics does not present a strength for local holding companies, members of the Bulgarian Industrial Capital Association. Demonstrating a deeper interest of senior managers in cultural perspectives in solving of business-related issues may increase to a greater extent organizational efficiency and effectiveness in this turbulent environment, because earlier identification of (and timely reflection on) key marker events provides adequate opportunities for planning and implementation of seamless change initiatives, regarding pursued strategies, enacted business models, demonstrated attitudes to employees and environment. That is why at least several recommendations may be given to senior management of these business organizations these are as follows:

• To enrich the content of their websites, going beyond the provision of general and succinct information about produced products and delivered services by the company by creating a clearly stated mission and vision statement in order to be able to explain to the target audience what is the desired situation for the company in the future and how it can get there or in what way it may happen.

• To discuss, reach agreement and officially declare official management principles of the firm in order to persuade the stakeholders in the preferred ways of achieving its goals.

• A balanced view on the interests of the constituencies that the company intends to consider and defend, should be constructed.

• The language that is used for the purpose of expressing professed corporate culture, should be clear and precise without using specific and exotic terms or loanwords.

• The list of official company values should be revised and its items should be carefully selected, based on discussions and reached agreement among the senior managers and even maybe the greater part of the personnel. These values should also be accompanied by clear and concise definitions for the convenience of the potential users.

• The arbitrary accumulation of content, regarding professed firm culture should be avoided, because of its acceptance as informational overload by different stakeholders.

• The maintained language versions of the professed culture content should be wisely chosen in congruence with the performed market and client segmentation.